# Spintronic temperature nanosensor based on the resonance response of a skyrmion-hosting magnetic tunnel junction

Michail Lianeris[1,2], Davi Rodrigues[1], Andrea Meo[1], Dimitris Kechrakos[2]
Anna Giordano[3], Mario Carpentieri[1], Giovanni Finocchio[3], and Riccardo Tomasello[1*]

[1]*Department of Electrical and Information Engineering, Politecnico di Bari, Via Edoardo Orabona, 4, 70126 Bari BA, Italy*

[2]*Physics Laboratory, Department of Education, School of Pedagogical and Technological Education (ASPETE), 15122 Athens, Greece*

[3]*Department of Mathematical and Computer Sciences, Physical Sciences and Earth Sciences, University of Messina, I-98166, Messina, Italy*

**Abstract**

The increasing need for efficient thermal management in nanoelectronics requires innovative thermal sensing solutions, as conventional sensors often exhibit nonlinear responses, low sensitivity, and complex calibration. We predict a temperature dependence in the response of existing skyrmion-based spintronic diodes and propose their use as nanoscale thermal sensors. These devices leverage magnetic skyrmions—topologically protected spin textures known for their robustness, nanoscale dimensions, and low-power dynamics. We demonstrate high thermal sensitivity with a linear temperature response over a wide range. This linearity, observed in both the amplitude and frequency of the skyrmion excitation, ensures redundancy that enables precise and reliable temperature measurement. In addition, the use of multilayer systems enhances the sensitivity and robustness of the device. These results provide a foundation for skyrmion-based caloritronic devices with promising applications in spintronic sensors, thermal management, nanoelectronics, and skyrmion-caloritronics

*Corresponding author: riccardo.tomasello@poliba.it

# 1. INTRODUCTION

Thermal sensors are essential for precise temperature monitoring in fields such as nanoelectronics, medical diagnostics, and industrial automation [1], [2], [3], [4], [5]. As devices shrink and power densities increase, effective thermal management becomes increasingly important. However, conventional sensors often suffer from non-linear responses, limited sensitivity at the nanoscale, and complex calibration [6], [7], [8], [9]. Linearity is particularly important because it simplifies calibration and improves accuracy and integration. These limitations hinder traditional sensors in nanoscale applications, highlighting the need for innovative, scalable and robust sensing technologies based on new materials and techniques[10], [11], [12], [13], [14].

Spintronics is an advanced field of electronics that exploits the intrinsic spin of electrons, in addition to their charge, to enable new functionalities and improved performance in sensing applications [15]. A cornerstone of spintronic technology is the magnetic tunnel junction (MTJ) that exploits the tunneling magnetoresistance (TMR) effect to convert the relative magnetization alignment of two ferromagnetic layers separated by an insulating barrier, typically magnesium oxide (MgO) [16], [17] into an electrical signal. MTJs have exhibited very good sensitivity combined with lower-power operation, which makes them apt for next-generation sensing application. Notably, MTJ networks have demonstrated excellent performance in magnetic field imaging [18] and energy harvesting[19].

An additional type of MTJ sensor can be based on the spin-torque diode (STD) effect [20]. This is a rectification phenomenon that involves the application of an ac input signal to the MTJ that is converted, via the TMR effect, into a dc output voltage. STDs operating in the linear resonance regime have been proposed as efficient nanoscale sensors for microwave signals[21] and mechanical vibrations [22], as well as electromagnetic energy harvesters[23], [24] when working in the broadband regime. More recently, STD MTJs have been proposed also for unconventional computing tasks[25].

The STD effect can also originate from the internal dynamics of magnetic solitons, such as vortices [26] and skyrmions [27], [28], [29]. In particular, magnetic skyrmions are nanoscale structures that exhibit rich quasi-particle dynamics[30]. Skyrmions emerge in systems with broken inversion symmetry, typically due to the presence of the Dzyaloshinskii-Moriya interaction (DMI). At interfaces, DMI arises from strong spin-orbit coupling in adjacent heavy metal layers. This interfacial DMI stabilizes Néel-type skyrmions, which are characterized by radially inward or outward magnetization around the core. Néel skyrmions exhibit ultra-low current and field-driven dynamics[31], [32], [33], [34].

The topological properties of skyrmions confer enhanced stability against thermal perturbations while enabling low-power excitation modes [11,12]. Skyrmions have demonstrated electrically detectable dynamics in response to applied currents [35], [36], voltage gates [28], temperature gradients [37] and strain [38]. Notably, their electrical readout can be further enhanced in multilayer systems[32], [39], [40]. These properties position skyrmions as promising candidates for low-power unconventional computing[41], [42], [43], [44], [45], [46], [47] and advanced sensing applications [27], [28], [32], [48]. More recently, the integration of MTJs with magnetic skyrmions has paved the way toward the design of more functional, sophisticated and highly tunable devices[32], [40], [49], [50].

The excitation modes of skyrmions can be both electrically driven and detected, and are highly sensitive to temperature variations, making them promising candidates for thermal sensing applications. Among these modes, the so-called breathing mode — characterized by radial oscillations of the skyrmion spin texture — has been extensively studied both theoretically and experimentally[29], [51], [52], [53], [54], [55], [56]. This mode can be driven by external stimuli such as spin-polarized currents, voltage gates, magnetic fields, and thermal fluctuations [27], [28], [57], [58]. The breathing frequency of skyrmions is in the GHz range [51], [59], [60], [61], and is highly tunable, depending on factors such as device geometry, material properties, and thermal conditions [37], [55], [62]. Due to their low power requirements and high operating frequencies, skyrmion excitations have already been proposed, and experimentally demonstrated, for microwave sensing[27], [28], [29]. Their inherent sensitivity to temperature further makes these platforms highly suitable for adaptation to precise thermal sensing applications. Previous proposals to use MTJs as thermal sensors have relied on thermally induced stochastic switching of the magnetization [63] or temperature-dependent variations in resistive behavior [64], [65]. However, the inherently probabilistic nature of magnetization dynamics in these schemes necessitates the use of large device arrays to ensure reliable and accurate measurements.

In the present work, we predict the temperature dependence of the STD in MTJs with a single skyrmion and propose a resonant spintronic temperature sensor. We perform a theoretical analysis, via systematic micromagnetic simulations and analytical calculations, on the effect of temperature on the resonant response of the skyrmion both in a MTJ with a single free layer and in a MTJ with a magnetic multilayer, since both systems have been proved to host skyrmions at room temperature [32], [40], [66]. We observed that both the skyrmion excitation frequency and the corresponding DC output voltage vary linearly with temperature, with sensitivities of about 2 µV/K for the single layer case and 0.7 µV/K for the multilayer case. We also compare the performance of a skyrmion-based device with that of a device containing uniform magnetization and show that the skyrmion-based configuration exhibits significantly greater sensitivity to temperature variations. Based on these promising results, we propose a realistic implementation of the device. Overall, our work paves the way for a reliable nanoscale temperature sensor suitable for wide-range temperature monitoring in nanoelectronics applications.

## 2. DEVICE AND MODELING
### 2.1. Device Description

We consider an MTJ consisting of a free layer (FL), an insulating barrier (typically MgO), and a reference layer [69],[70],[71],[72]. We only simulate and study the magnetization dynamics of the FL. Experimental skyrmion-hosting MTJs have used so far two designs of the FL: (i) a single FL coupled to a heavy metal (HM) layer, essential to provide a strong-enough interfacial DMI (IDMI)[66]; (ii) a FL coupled with a magnetic multilayer which acts as a skyrmion injector [32], [40], [67]. Hereafter, we will refer to design (i) as single FL MTJ, and to design (ii) as multilayer MTJ. We simulate cylindrical MTJs with diameter of 300nm, a FL thickness of 1nm for the single-layer FL MTJ, shown in Figure 1(a), and a multilayer FM consisting of five repetitions of a 1 nm thick ferromagnetic layer separated by 1 nm thick HM layers, as sketched in Figure 1(b).

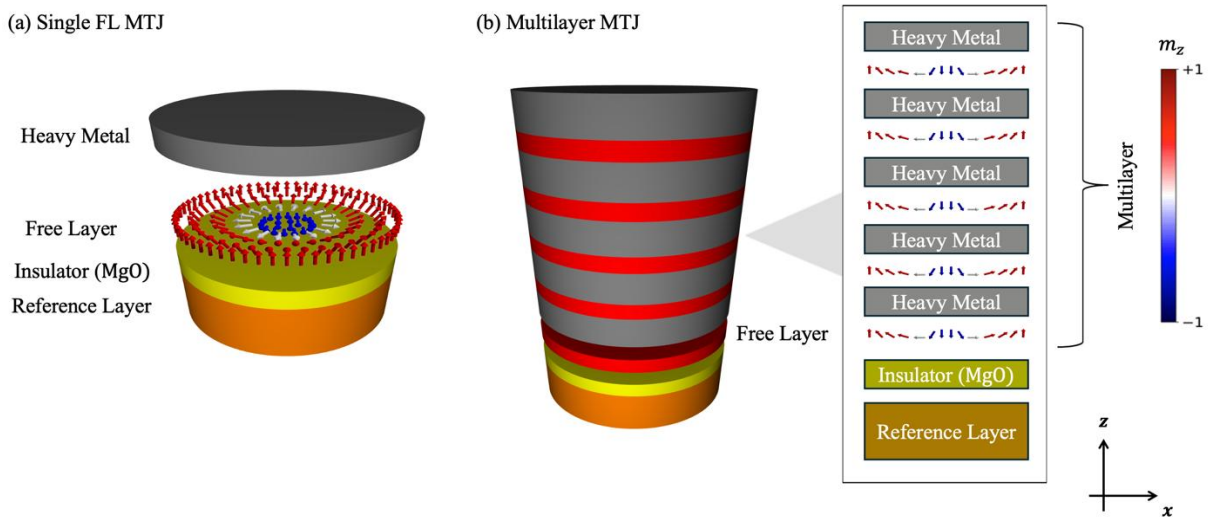

Figure 1: (a) Sketch of the single FL MTJ composed of a bottom fixed layer, an insulating layer (MgO) layer, a free layer and a heavy metal layer on top. The free magnetic layers host room-temperature Néel skyrmions. (b) Sketch of the multilayer MTJ, composed of a bottom fixed layer, an insulating layer (MgO) layer, a free layer and a multilayer stack of skyrmion-hosting layers, as depicted on the right. The color palette represents the out-of-plane component of the magnetization. The reference system is also shown.

## 2.2. Micromagnetic Model

The micromagnetic simulations are performed using PETASPIN, a state-of-the-art full micromagnetic solver. This tool utilizes the Adams-Bashforth time solver scheme for the numerical integration of the Landau-Lifshitz-Gilbert (LLG) equation [37], [68]

$$\frac{\partial m}{\partial \tau} = -(m \times h_{eff}) + \alpha_G \left(m \times \frac{\partial m}{\partial \tau}\right) \quad (1)$$

where $\mathbf{m} = \mathbf{M}/M_s$ is the normalized magnetization vector, $\alpha_G$ is the Gilbert damping and $\tau = \gamma_0 t M_s$ is the dimensionless time, with $\gamma_0$ being the gyromagnetic ratio and $M_s$ the saturation magnetization. $\mathbf{h_{eff}}$ is the normalized effective field, which includes the exchange (A), interfacial DMI (D)[69], uniaxial anisotropy ($K_u$), magnetostatic and external field (H) contributions.

We excite the MTJ FL via the spin-transfer torque ($\tau_{STT}$) induced by a perpendicularly polarized ac current. This term is added to the right-hand side of Eq. (1):

$$\tau_{STT} = -\frac{g\mu_B P j_0}{\gamma_0 e M_s^2 t_{FM}}[\mathbf{m} \times (\mathbf{m} \times \hat{\mathbf{p}})] \quad (2)$$

with g being the Landé factor, $\mu_B$ the Bohr magneton, e the electron charge, $t_{FM}$ the thickness of the MTJ FL, P the polarization coefficient of the injected electrical current, $j_0$ is the electric ac current density and $\hat{\mathbf{p}}$ is the direction of the spin- polarization, directed along the positive z-axis.

Finite temperature effects are included in our micromagnetic model via well-established temperature-scaling relations of the material parameters[28], [37]. In particular, the change in saturation magnetization with temperature as $M_s(T) = M_s(0)(1 - (T/T_c)^\delta)$ with $M_s(0)$ being the saturation magnetization of the ferromagnet at zero temperature, $\delta = 1.5$ and $T_c = 1120$ K is the Curie temperature [70], [71]. We scale the values of the rest of the material parameters as:

$$A(T) = A(0)m(T)^\alpha, \quad D(T) = D(0)m(T)^\beta,$$
$$K_u(T) = K_u(0)m(T)^\gamma \quad (3)$$

with $m(T) = M_s(T)/M_s(0)$. We perform micromagnetic simulations in the range 0 to 400 K, using the zero temperature parameters[29], [72], shown in Table 1.

For the single FL MTJ, we use the scaling exponents α = β = 1.5, γ = 3.0[72] and for the multilayer MTJ, we use α = 1.7, β = 2.0, γ = 2.5[73], [74]. In all cases, we use a discretization cell size of 3 x 3 x 1 nm³. We neglect the temperature dependence of the damping, as it increases only by 5% in the range 0 – 400K [75], [76], [77]. For the single FL MTJ, we consider the thin film approximation, assuming uniform magnetization across the layer thickness. In this approximation, the average effect of the magnetostatic interactions is treated as an effective modification of the perpendicular anisotropy. In contrast, for the multilayer MTJ, we perform a full calculation of the magnetostatic interactions that captures the resulting variations in the skyrmion profile across the different layers [72].

We obtain the rectified output voltage ($V_{dc}$) as the time average of the voltage V(t) across the MTJ. In particular, V(t) = I(t)R(t), with I(t) = A$j_0$sin(ωt) the exciting ac current, A the cross-section area of the device and $j_0$ the current density. Also, R(t) = $R_p$ + 0.5($R_{ap}$ − $R_p$)(1 − $m_z$(t)) is the magnetization-dependent resistance of the MTJ due to the TMR effect, with $R_{p(ap)}$ the value of the MTJ resistance in the parallel (antiparallel) configuration. The values of $R_{ap}$ = 1.5 kΩ and $R_p$ = 1.0 kΩ are taken from previous works [27], [28].

Table 1: Summary of the micromagnetic parameters at zero temperature used in the simulations [29,72].

| Parameters | Single FL MTJ | Multilayer MTJ |
|---|---|---|
| A(0) (pJ/m) | 20 | 15 |
| $K_u$(0)(MJ/m³) | 0.75 | 0.85 |
| D(0) (mJ/m²) | 3.0 | 3.0 |
| H (mT) | 0 | 10-70 |
| $M_s$(0) (kA/m) | 600 | 900 |
| $α_G$(0) | 0.1 | 0.1 |

We obtain the rectified output voltage ($V_{dc}$) across the STD as the time average of the voltage V(t) across the MTJ. $V(t) = I(t)R(t)$, with $I(t) = Aj_0\sin(\omega t)$ the exciting ac current, A the cross-section area of the device and $j_0$ the current density. $R(t) = R_p + 0.5(R_{ap} - R_p)(1 - m_z(t))$ is the magnetization-dependent resistance of the MTJ due to the TMR effect, with $R_{p(ap)}$ the value of the MTJ resistance in the parallel (antiparallel) configuration. The values of $R_{ap} = 1.5$ k$\Omega$ and $R_p = 1.0$ k$\Omega$ are taken from previous works [27,28].

### 2.3. Analytical Model

The breathing mode of a magnetic skyrmion refers to a radial oscillation of its size, characterized by time-dependent variations in both the skyrmion radius R and the azimuthal angle φ, the latter also known as the helicity. These variables serve as collective coordinates of a soft excitation mode and fully describe the breathing dynamics within a reduced analytical model [55].

In the limit of large skyrmions, where the radius R is significantly larger than the domain wall width $\Delta = \sqrt{A/K}$, the total energy of a skyrmion can be approximated by the following expression:

$$E_{Sk}(R) = \sqrt{\frac{A^3}{K}}\left((1 - 2g\cos(\varphi))\left(\frac{R}{\Delta}\right) + \frac{2\Delta}{R}\right) \quad (4)$$

Here $K = K_u - 0.5\mu_0 M_s^2$ is the effective anisotropy energy density and $g = \pi D/(4\sqrt{AK})$ is a coupling constant. Minimizing the energy with respect to R yields the equilibrium skyrmion radius:

$$R_{SK} = \frac{1}{\sqrt{2(1-g)}} \quad (5)$$

Since R and φ are dynamical variables associated with the breathing mode, they satisfy a Poisson bracket of the form: $[\varphi, R^2] = k\gamma_0/(t_{FM}M_s)$, where k is a fitting parameter that depends on the precise skyrmion profile. Applying the Hamiltonian formalism to this reduced model, the breathing mode frequency $\omega_0$ for small-amplitude oscillations can be derived as [55], [78]:

$$\omega_F = \frac{k\pi\gamma_0 K}{t_{FM}M_s}\frac{1}{R_{SK}\sqrt{1+R_{SK}^2}} \quad (6)$$

This expression predicts the frequency of skyrmion breathing modes in thin films accounting for the interplay of exchange interaction, magnetic anisotropy, dipolar fields and DMI.

## 3. RESULTS

### 3.1. Temperature dependence of skyrmion size

First, we verify that a single skyrmion can be stabilized in the FL of both types of MTJs around room temperature. Therefore, we analyze the dependence of the skyrmion stability and equilibrium size under different temperatures, obtained by using Eq. (3), and external fields. We place an initial Néel skyrmion in the FL of the MTJ and subsequently relax the system.

Figure 2(a) shows that, in the single FL MTJ, the Néel skyrmion is stabilized within a temperature range between 200K and 400K. We observe a pronounced increase in the skyrmion radius as the material parameters change with increasing temperature, as found in our previous work [72]. This behavior is attributed to the temperature-induced variation of material parameters, which in turn affects the energy of the skyrmion configuration. We also note that the behavior of the radius as a function of the temperature is in good agreement with the analytical predictions of Eq. (5), when considering the temperature scaling factors for the magnetic parameters.

In the multilayer MTJ we obtain pure Néel skyrmions[79] and the radius remains almost constant over the entire temperature range from 200 to 400 K, as shown in Figure 2(b). We attribute this behavior to two competing factors: On the one hand, the magnetostatic field favors an increase of the skyrmion size, on the other, the faster decrease of DMI with temperature with respect to the exchange energy ($\beta > \alpha$) favors a reduction of the skyrmion size.

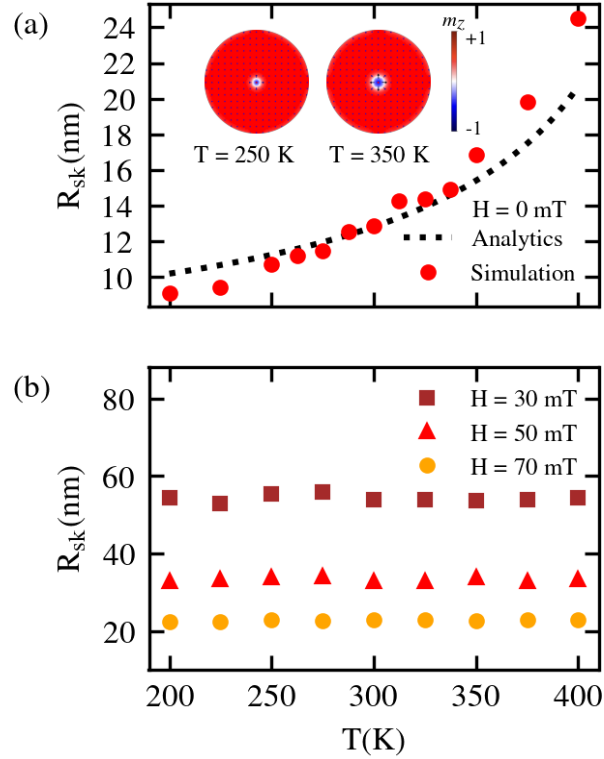

Figure 2: Temperature dependence of the Néel skyrmion radius ($R_{sk}$) in (a) single FL MTJ and (b) in the FL of the multilayer MTJ. In (a) $R_{sk}$ increases with temperature, indicating a direct thermal influence on skyrmion size. In (b) $R_{sk}$ remains constant with temperature, highlighting the temperature-insensitive behavior of skyrmion size in multilayer structures.

### 3.2. Temperature sensor based on the single FL MTJ

We analyze the STD effect of a Néel skyrmion in a single FL MTJ under an ac current with density $j_0 = 1.0 \text{ MA/cm}^2$ as a function of temperature. The goal here is to establish a link between the feature of the skyrmion STD response (e.g. resonance frequency and/or $V_{dc}$) and the temperature to be measured. For the STD effect, we consider that the variation of the device resistance is related to the ac-driven breathing mode of the skyrmion[27], [28].

We sweep the frequency of the applied ac current[80] and obtain the resonance curves shown in Figure 3 for different temperatures, from 250 to 350 K. We observe a strong red shift, i.e., a shift towards lower frequency, of the resonant frequency with an increasing temperature, ranging from 9 GHz at 250 K to 5 GHz at 350 K. In addition, the magnitude of the peak value of the $V_{dc}$ at the resonance frequency shows a net temperature dependence, which can be attributed to the strong temperature dependence of the skyrmion size in the single FL MTJ (see Figure 2(a)).

The characteristic breathing frequency and amplitude are related to the equilibrium radius of the skyrmion (see Eq. (6)), such that larger skyrmions present a lower resonant frequency and larger oscillation amplitude at the resonance [55].

Next, we extract the characteristics of the resonant behavior of the rectified voltage. Figure 4(a) shows the magnitude of the peak value of $V_{dc}$ as a function of temperature for $j_0 = 1.0 \, \text{MA/cm}^2$. A clear linear behaviour is observed in the temperature range around room temperature. For the sake of comparison, we also show in Figure 4(a) the maximum rectified voltage obtained for the same device, but without the presence of a skyrmion in the FL of the MTJ, i.e., uniform perpendicular magnetization. Clearly, the skyrmion-free device exhibits a much lower temperature sensitivity, that is quantified by the much lower slope of the data. This comparison highlights the superior efficiency of the skyrmion-MTJ device compared to the uniform-layer MTJ for the temperature detection. Figure 4(b) shows the resonant frequency of the skyrmion as a function of temperature, which shows good agreement with the analytical prediction in Eq. (6), using a fitting parameter of $k = 0.25$. The linear temperature dependence of both the resonant frequency ($\omega_F$) and the maximum $V_{dc}$ suggests two independent but complementary sensing mechanisms - monitoring the resonant amplitude and tracking the resonant frequency. Together, these methods provide a reliable and robust approach to accurate temperature measurement. It is important to note that, although the frequency of the breathing mode generally scales with the inverse square of the skyrmion radius, the radius itself varies with temperature in such a way that the frequency has an approximately linear dependence over the temperature range considered. For the single FL MTJ, the Taylor expansion of the analytical breathing mode frequency $\omega_0(T)$ as a function of temperature yields $\omega_0(T) \cong 6.5 \times 10^9 - 4.93 \times 10^7 (T - 300) - 1.6652 \times 10^5 (T - 300)^2 + 48(T - 300)^3 + O((T - 300)^4)$. It demonstrates that higher order terms beyond the linear component are significantly smaller, confirming that the temperature dependence of the frequency can be well approximated as linear in this range.

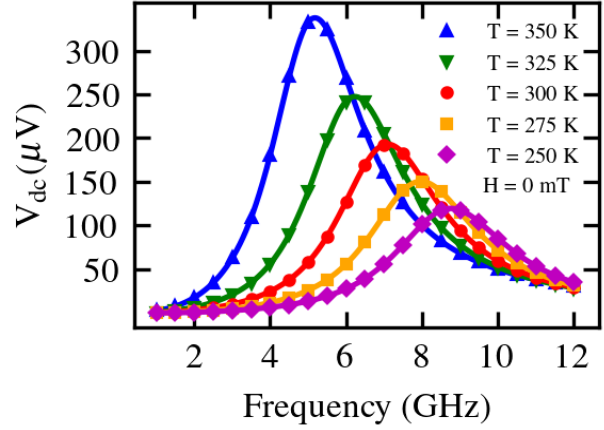

Figure 3: Rectified voltage ($V_{dc}$) of the single FL MTJ as a function of temperature for a current density amplitude $j_0 = 1.0 \text{ MA/cm}^2$. The resonant frequency exhibits a red shift with temperature and the maximum value of the rectified voltage an increase with temperature.

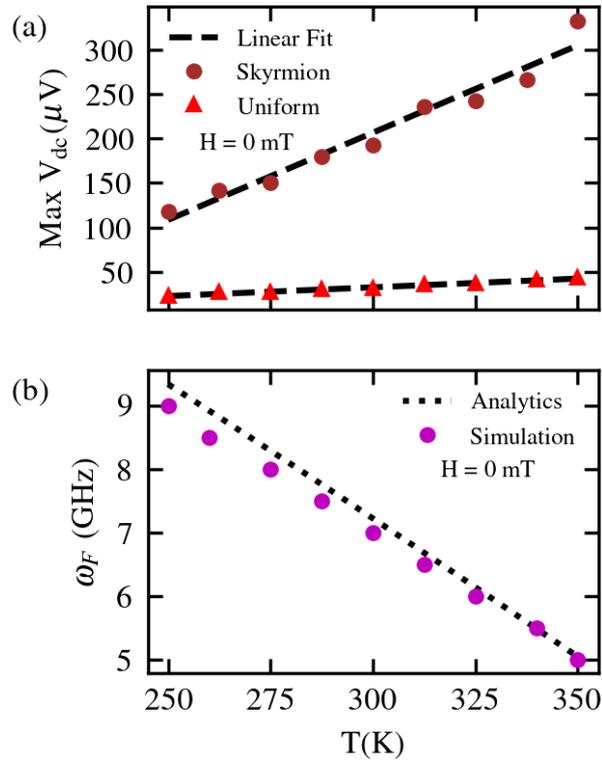

Figure 4: Temperature dependence of (a) the maximum rectified voltage $V_{dc}$ and (b) resonant frequency ($\omega_F$) the single FL MTJ for a current density amplitude $j_0 = 1.0 \text{ MA/cm}^2$. For comparison, the case of uniform magnetization is also shown. Linese in (a) are fits to a line, while in (b) describe $\omega_0(T)$, Eq.(6) with fitting parameter $k = 0.25$.

### 3.3. Temperature sensor based on the multilayer MTJ

Next, we investigate the STD effect in the multilayer MTJ. The strategy of using such a system instead of a single FL MTJ has been already and successfully adopted in experiments [32], [40], [67]. Figure 5 shows the frequency dependence of $V_{dc}$ for different temperatures around room temperature. The multilayer MTJ produces sharper resonant peaks characterized by narrower bandwidth ($\Delta f \sim 1GHz$) compared to the single-layer configuration ($\Delta f \sim 3GHz$) seen in Figure 3. We attribute the narrower $\Delta f$ to the enhanced thermal stability of skyrmions in the multilayer structure, which allows for more accurate sensing of both resonant frequency and peak amplitude.

In Figure 6 we present the temperature dependence of the maximum $V_{dc}$ and $\omega_F$ for the multilayer MTJ. The data show a linear behavior in both cases. Compared to the single FL MTJ, $\omega_F$ exhibits a smaller slope indicating lower sensitivity in a potential sensing application based on the temperature dependence of the resonant frequency of a multilayer configuration. Nonetheless, the reduced slope of the $\omega_F(T)$ is compensated by the narrower bandwidth of the resonance curves (see Figure 5) that provide a higher accuracy in the determination of the resonant frequency, reinforcing overall the superior performance of the multilayer configuration relative to the single layer one.

The proposed devices have a diameter of 300 nm and a height of less than 40 nm, depending on the specific configuration. For the single FL MTJ, we achieved a sensitivity of about 2 µV/K, while the multilayer MTJ showed a sensitivity of about 0.7 µV/K. Although these values are modest when compared to conventional commercial thermal sensors, there is considerable potential for enhancement through systematic optimization — including careful material selection and geometric refinement. In particular, the sensitivity of the multilayer MTJ could be increased by operating at zero field, a scenario experimentally achievable [29].

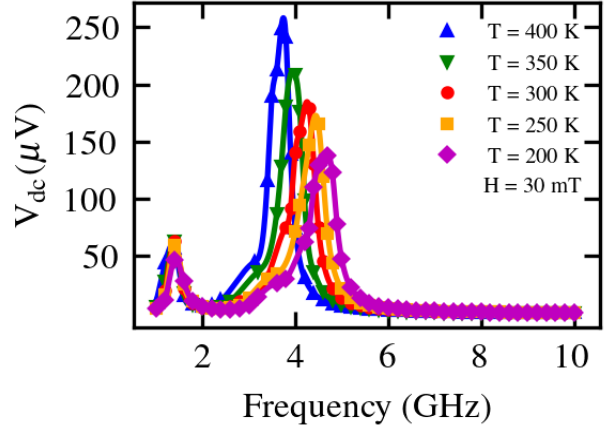

Figure 5: Rectified voltage ($V_{dc}$) of the multilayer MTJ as a function of temperature for a current density amplitude $j_0 = 1.0$ MA/cm$^2$. The resonant frequency exhibits a red shift with temperature and the maximum rectified voltage increases with temperature.

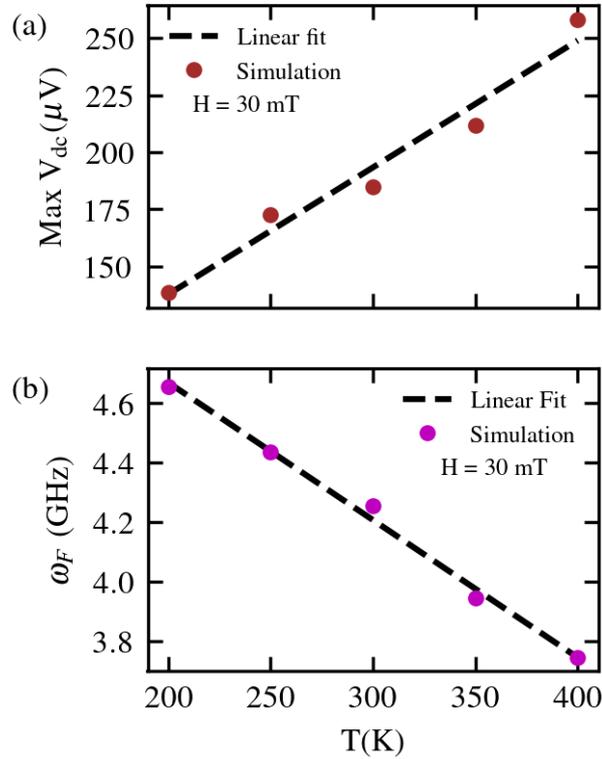

Figure 6: Temperature dependence of the maximum rectified voltage $V_{dc}$ and resonant frequency $\omega_F$ of the multilayer MTJ composed of 5 FM layers with a current density amplitude $j_0 = 1.0$ MA/cm$^2$. $\omega_F$ decreases linearly with temperature from 4.7 GHz at 200 K to 3.8 GHz at 400 K. Proposed experimental implementation of the temperature sensor.

## 4. DISCUSSION

Given the compatibility of the proposed device with CMOS fabrication processes, these devices can be directly integrated into nanoscale systems, either at specific points or distributed throughout the architecture, enabling precise local temperature monitoring[64]. Figure 7 illustrates a conceptual example of their potential application for temperature sensing within computer processing units (CPUs).

Beyond the heat flow from the CPU to the proposed thermal sensor, we estimate a temperature rise ($\Delta T$) in the single FL MTJ sensor caused by Joule heating from the applied ac current. The power dissipated in the device is given by $P = \langle I^2(t) \cdot R(t) \rangle$, where $\langle \cdot \rangle$ denotes the time-averaged value and $R(t)$ is the time-dependent electrical resistance of the device (see Section 2.2). According to Fourier's law, the steady state temperature rise can be expressed as $P = G \cdot \Delta T$, where G is the thermal conductance of the device.

The thermal conductance is determined by the device geometry and material properties and can be approximated by $G = \kappa \cdot (A/d)$, where A is the cross-sectional area, d is the MgO thickness, and $\kappa$ is the thermal conductivity. In this case, the main contribution to $\kappa$ arises from the metal–MgO interface, specifically for CoFeB–MgO, with reported values of $\kappa \approx 0.15\text{-}4$ $\text{Wm}^{-1}\text{K}$[81], [82]. Based on our micromagnetic simulations, the power dissipated in the sensor is approximately $P \approx 8 - 12\ \mu\text{W}$ across the applied ac frequency range of 1–12 GHz. For a device with a 300 nm diameter and 1 nm MgO thickness, this yields a temperature rise of approximately $\Delta T \approx 0.3 - 2.0$ K, which is negligible in comparison to the target measurement range of $300 - 350$ K. Thus, Joule heating is expected to have a minute effect on the accuracy of temperature sensing of the proposed device. We also notice that the sensitivity reduces as the FL thickness increases, while the power consumption is not significantly affected.

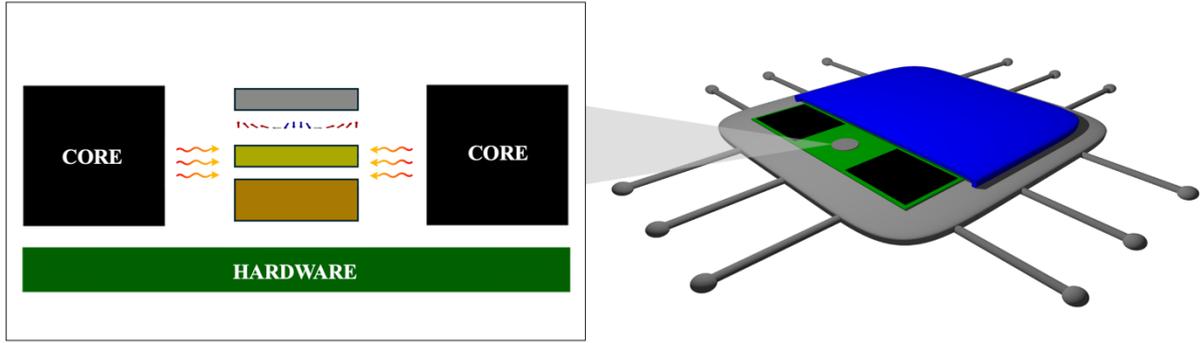

Figure 7: Sketch of an MTJ skyrmion sensor integrated into a CPU. The sensor is placed between two cores to measure local temperature fluctuations. The MTJ sensor is illustrated by (a) a cross-section and (b) its placement within the CPU.

## 5. SUMMARY AND CONCLUSIONS

In this work, we predict a temperature dependence of the STD in existing MTJs with a single skyrmion and propose a skyrmion-based temperature sensor that leverages the low-power dynamics of magnetic skyrmions and the scalability, CMOS compatibility, and robustness of MTJs for efficient and reliable nanoscale temperature measurement. The sensor exploits the temperature-dependent breathing mode of skyrmions, combined with the spin-torque diode effect, to provide two independent and complementary temperature metrics: the maximum rectified voltage ($V_{dc}$), and the resonant frequency ($\omega_F$).

Numerical and analytical results show a clear linear response for both metrics in the temperature range of 200 K to 400 K, enabling accurate and robust measurements without the need for complex calibration procedures. The temperature dependence of $V_{dc}$ and $\omega_F$ has been systematically characterized for both the single FL MTJ and the multilayer MTJ. These behaviors are directly related to the thermal sensitivity of the magnetic material parameters and the intrinsic tunability of the magnetic skyrmions.

Using realistic experimental parameters, the single-layer MTJ sensor showed a sensitivity of about 2 µV/K, while the multilayer counterpart showed a lower sensitivity of about 0.7 µV/K. Inclusion of thermal fluctuations in our micromagnetic study via a stochastic LLG approach, has shown that the resonance frequency is not modified while the linewidth of the resonance increases and the peak value of Vdc reduces. These results suggest that thermal fluctuations slightly reduce the device's sensitivity.

The use of thermal fluctuations instead of scaling relations of the material parameters can lead to a further reduction of the sensitivity. Although these values are low compared to the state-of-the-art sensors, this is a first proposal based on STD and topological textures and further optimization of the device may improve the performance. Possible strategies include: the use of a FL with a smaller damping (our estimation with $\alpha_G$=0.021 gives a sensitivity of about 10 µV/K); simultaneous application of voltage-controlled magnetocrystalline anisotropy (VCMA) [66]; active STDs where the additional dc current can excite the injection locking phenomenon, as already experimentally demonstrated for MTJs with a uniform FL [83]. The multilayer design offers significant advantages, including improved skyrmion stability and sharply defined resonance peaks - with a bandwidth three times narrower than that of the single-layer system - resulting in improved measurement accuracy. In addition, the proposed sensor operates with low power consumption (~10 µW) and minimal Joule heating, reinforcing its suitability for energy-efficient and thermally sensitive applications. In particular, the multilayer architecture shows strong potential for further optimization, especially considering recent advances in skyrmion stability and detection in multilayer systems[29], [66].

Overall, these results provide a promising solution for the development of thermally controlled nanotechnology, particularly for GHz frequency applications where precision, adaptability and thermal management are essential. This work also paves the way for future innovations in skyrmion-based sensing and energy-efficient spintronics technologies.

## ACKNOWLEDGMENTS

This work was supported by the projects "SKYrmion-based magnetic tunnel junction to design a temperature SENSor—SkySens", funded by the Italian Ministry of Research, and by the project number 101070287—SWAN-on-chip— HORIZON-CL4- 2021-DIGITAL EMERGING-01. Support from the project PE0000021, "Network 4 Energy Sustainable Transition - NEST", funded by the European Union - Next Generation EU, under the National Recovery and Resilience Plan (NRRP), Mission 4 Component 2 Investment 1.3 - Call for Tender No. 1561 dated 11.10.2022 of the Italian MUR (CUP C93C22005230007). The authors are with the PETASPIN team and thank the PETASPIN association (www. petaspin.com). ML acknowledges hospitality in the Physics Laboratory, Department of Education, School of Pedagogical and Technological Education (ASPETE), where part of this work was performed.

# REFERENCES

[1] X. Sotvoldiev, D. Tukxtasinov, S. Zokirov, S. Toxirova, M. Abdullayeva, and A. Muhammadjonov, "Review and analysis of methods of automation of temperature measurement process," *E3S Web of Conferences*, vol. 592, p. 03024, Nov. 2024, doi: 10.1051/e3sconf/202459203024.

[2] A. Balčytis, M. Ryu, S. Juodkazis, and J. Morikawa, "Micro-thermocouple on nano-membrane: thermometer for nanoscale measurements," *Sci Rep*, vol. 8, no. 1, p. 6324, Apr. 2018, doi: 10.1038/s41598-018-24583-w.

[3] S. R. Madhvapathy *et al.*, "Advanced thermal sensing techniques for characterizing the physical properties of skin," *Appl Phys Rev*, vol. 9, no. 4, Dec. 2022, doi: 10.1063/5.0095157.

[4] M. Javaid, A. Haleem, R. P. Singh, S. Rab, and R. Suman, "Significance of sensors for industry 4.0: Roles, capabilities, and applications," *Sensors International*, vol. 2, p. 100110, 2021, doi: 10.1016/j.sintl.2021.100110.

[5] F. Reverter, "A Tutorial on Thermal Sensors in the 200th Anniversary of the Seebeck Effect," *IEEE Sens J*, vol. 21, no. 20, pp. 22122–22132, Oct. 2021, doi: 10.1109/JSEN.2021.3105546.

[6] O. Kuldashov, O. Rayimdjanova, B. Djalilov, S. Ergashev, S. Toxirova, and A. Muhammadjonov, "Stabilization of parameters of optoelectronic devices on semiconductor emitters," *E3S Web of Conferences*, vol. 508, p. 01001, Apr. 2024, doi: 10.1051/e3sconf/202450801001.

[7] N. V. S. R. Nalakurthi *et al.*, "Challenges and Opportunities in Calibrating Low-Cost Environmental Sensors," *Sensors*, vol. 24, no. 11, p. 3650, Jun. 2024, doi: 10.3390/s24113650.

[8] F. Chi *et al.*, "Multimodal temperature sensing using $Zn_2GeO_4:Mn^{2+}$ phosphor as highly sensitive luminescent thermometer," *Sens Actuators B Chem*, vol. 296, p. 126640, Oct. 2019, doi: 10.1016/j.snb.2019.126640.

[9] Y. Han *et al.*, "Design of Hetero-Nanostructures on $MoS_2$ Nanosheets To Boost $NO_2$ Room-Temperature Sensing," *ACS Appl Mater Interfaces*, vol. 10, no. 26, pp. 22640–22649, Jul. 2018, doi: 10.1021/acsami.8b05811.

[10] M. Asheghi, K. Kurabayashi, R. Kasnavi, and K. E. Goodson, "Thermal conduction in doped single-crystal silicon films," *J Appl Phys*, vol. 91, no. 8, pp. 5079–5088, Apr. 2002, doi: 10.1063/1.1458057.

[11] H. Chen *et al.*, "High toughness multifunctional organic hydrogels for flexible strain and temperature sensor," *J Mater Chem A Mater*, vol. 9, no. 40, pp. 23243–23255, 2021, doi: 10.1039/D1TA07127K.

[12] J. Cai, M. Du, and Z. Li, "Flexible Temperature Sensors Constructed with Fiber Materials," *Adv Mater Technol*, vol. 7, no. 7, Jul. 2022, doi: 10.1002/admt.202101182.

[13] E. Song *et al.*, "$Mn^{2+}$-activated dual-wavelength emitting materials toward wearable optical fibre temperature sensor," *Nat Commun*, vol. 13, no. 1, p. 2166, Apr. 2022, doi: 10.1038/s41467-022-29881-6.

[14] Y. Han *et al.*, "Design of Hetero-Nanostructures on $MoS_2$ Nanosheets To Boost $NO_2$ Room-Temperature Sensing," *ACS Appl Mater Interfaces*, vol. 10, no. 26, pp. 22640–22649, Jul. 2018, doi: 10.1021/acsami.8b05811.

[15] B. Dieny *et al.*, "Opportunities and challenges for spintronics in the microelectronics industry," *Nat Electron*, vol. 3, no. 8, pp. 446–459, Aug. 2020, doi: 10.1038/s41928-020-0461-5.

[16] S. Ikeda *et al.*, "A perpendicular-anisotropy CoFeB–MgO magnetic tunnel junction," *Nat Mater*, vol. 9, no. 9, pp. 721–724, Sep. 2010, doi: 10.1038/nmat2804.

[17] Shinji Yuasa *et al.*, "Giant tunneling magnetoresistance in MgO-based magnetic tunnel junctions and its industrial applications," in *2006 IEEE Nanotechnology Materials and Devices Conference*, IEEE, Oct. 2006, pp. 186–187. doi: 10.1109/NMDC.2006.4388737.

[18] C. Ye, Y. Wang, and Y. Tao, "High-Density Large-Scale TMR Sensor Array for Magnetic Field Imaging," *IEEE Trans Instrum Meas*, vol. 68, no. 7, pp. 2594–2601, Jul. 2019, doi: 10.1109/TIM.2018.2866299.

[19] R. Sharma *et al.*, "Electrically connected spin-torque oscillators array for 2.4 GHz WiFi band transmission and energy harvesting," *Nat Commun*, vol. 12, no. 1, p. 2924, May 2021, doi: 10.1038/s41467-021-23181-1.

[20] G. Finocchio *et al.*, "Perspectives on spintronic diodes," *Appl Phys Lett*, vol. 118, no. 16, Apr. 2021, doi: 10.1063/5.0048947.

[21] A. A. Tulapurkar *et al.*, "Spin-torque diode effect in magnetic tunnel junctions," *Nature*, vol. 438, no. 7066, pp. 339–342, Nov. 2005, doi: 10.1038/nature04207.

[22] A. Meo *et al.*, "Magnetomechanical Accelerometer Based on Magnetic Tunnel Junctions," *Phys Rev Appl*, vol. 20, no. 3, p. 034003, Sep. 2023, doi: 10.1103/PhysRevApplied.20.034003.

[23] R. Sharma *et al.*, "Electrically connected spin-torque oscillators array for 2.4 GHz WiFi band transmission and energy harvesting," *Nat Commun*, vol. 12, no. 1, p. 2924, May 2021, doi: 10.1038/s41467-021-23181-1.

[24] B. Fang *et al.*, "Experimental Demonstration of Spintronic Broadband Microwave Detectors and Their Capability for Powering Nanodevices," *Phys Rev Appl*, vol. 11, no. 1, p. 014022, Jan. 2019, doi: 10.1103/PhysRevApplied.11.014022.

[25] L. Mazza *et al.*, "Computing with Injection-Locked Spintronic Diodes," *Phys Rev Appl*, vol. 17, no. 1, p. 014045, Jan. 2022, doi: 10.1103/PhysRevApplied.17.014045.

[26] A. S. Jenkins *et al.*, "Spin-torque resonant expulsion of the vortex core for an efficient radiofrequency detection scheme," *Nat Nanotechnol*, vol. 11, no. 4, pp. 360–364, Apr. 2016, doi: 10.1038/nnano.2015.295.

[27] G. Finocchio *et al.*, "Skyrmion based microwave detectors and harvesting," *Appl Phys Lett*, vol. 107, no. 26, Dec. 2015, doi: 10.1063/1.4938539.

[28] D. R. Rodrigues, R. Tomasello, G. Siracusano, M. Carpentieri, and G. Finocchio, "Ultra-sensitive voltage-controlled skyrmion-based spintronic diode," *Nanotechnology*, vol. 34, no. 37, p. 375202, Sep. 2023, doi: 10.1088/1361-6528/acdad6.

[29] B. Fang *et al.*, "Topological spin-torque diode effect in skyrmion-based magnetic tunnel junctions," p. arXiv:2405.10753, May 2024, [Online]. Available: http://arxiv.org/abs/2405.10753

[30] K. Everschor-Sitte, J. Masell, R. M. Reeve, and M. Kläui, "Perspective: Magnetic skyrmions—Overview of recent progress in an active research field," *J Appl Phys*, vol. 124, no. 24, Dec. 2018, doi: 10.1063/1.5048972.

[31] A. Soumyanarayanan *et al.*, "Tunable room-temperature magnetic skyrmions in Ir/Fe/Co/Pt multilayers," *Nat Mater*, vol. 16, no. 9, pp. 898–904, Sep. 2017, doi: 10.1038/nmat4934.

[32] Y. Guang *et al.*, "Electrical Detection of Magnetic Skyrmions in a Magnetic Tunnel Junction," *Adv Electron Mater*, vol. 9, no. 1, p. 2200570, Jan. 2023, doi: 10.1002/aelm.202200570.

[33] S. D. Pollard, J. A. Garlow, J. Yu, Z. Wang, Y. Zhu, and H. Yang, "Observation of stable Néel skyrmions in cobalt/palladium multilayers with Lorentz transmission electron

microscopy," *Nat Commun*, vol. 8, no. 1, p. 14761, Mar. 2017, doi: 10.1038/ncomms14761.

[34] Y. Tokura and N. Kanazawa, "Magnetic Skyrmion Materials," *Chem Rev*, vol. 121, no. 5, pp. 2857–2897, Mar. 2021, doi: 10.1021/acs.chemrev.0c00297.

[35] V. T. Pham *et al.*, "Fast current-induced skyrmion motion in synthetic antiferromagnets," *Science (1979)*, vol. 384, no. 6693, pp. 307–312, Apr. 2024, doi: 10.1126/science.add5751.

[36] S. Woo *et al.*, "Current-driven dynamics and inhibition of the skyrmion Hall effect of ferrimagnetic skyrmions in GdFeCo films," *Nat Commun*, vol. 9, no. 1, p. 959, Mar. 2018, doi: 10.1038/s41467-018-03378-7.

[37] E. Raimondo *et al.*, "Temperature-Gradient-Driven Magnetic Skyrmion Motion," *Phys Rev Appl*, vol. 18, no. 2, p. 024062, Aug. 2022, doi: 10.1103/PhysRevApplied.18.024062.

[38] Y. Yang *et al.*, "Acoustic-driven magnetic skyrmion motion," *Nat Commun*, vol. 15, no. 1, p. 1018, Feb. 2024, doi: 10.1038/s41467-024-45316-w.

[39] E. Raimondo *et al.*, "Micromagnetic study of the frequency response of skyrmions in magnetic multilayers," in *2024 IEEE 24th International Conference on Nanotechnology (NANO)*, IEEE, Jul. 2024, pp. 191–195. doi: 10.1109/NANO61778.2024.10628854.

[40] S. Chen *et al.*, "All-electrical skyrmionic magnetic tunnel junction," *Nature*, vol. 627, no. 8004, pp. 522–527, Mar. 2024, doi: 10.1038/s41586-024-07131-7.

[41] D. Prychynenko *et al.*, "Magnetic Skyrmion as a Nonlinear Resistive Element: A Potential Building Block for Reservoir Computing," *Phys Rev Appl*, vol. 9, no. 1, p. 14034, 2018, doi: 10.1103/PhysRevApplied.9.014034.

[42] G. Bourianoff, D. Pinna, M. Sitte, and K. Everschor-Sitte, "Potential implementation of reservoir computing models based on magnetic skyrmions," *AIP Adv*, vol. 8, no. 5, p. 055602, May 2018, doi: 10.1063/1.5006918.

[43] D. Pinna, G. Bourianoff, and K. Everschor-Sitte, "Reservoir Computing with Random Skyrmion Textures," *Phys Rev Appl*, vol. 14, no. 5, p. 054020, Nov. 2020, doi: 10.1103/PhysRevApplied.14.054020.

[44] Y. Huang, W. Kang, X. Zhang, Y. Zhou, and W. Zhao, "Magnetic skyrmion-based synaptic devices," *Nanotechnology*, vol. 28, no. 8, p. 08LT02, Feb. 2017, doi: 10.1088/1361-6528/aa5838.

[45] K. M. Song *et al.*, "Skyrmion-based artificial synapses for neuromorphic computing," *Nat Electron*, vol. 3, no. 3, pp. 148–155, Mar. 2020, doi: 10.1038/s41928-020-0385-0.

[46] N. Sisodia, J. Pelloux-Prayer, L. D. Buda-Prejbeanu, L. Anghel, G. Gaudin, and O. Boulle, "Programmable Skyrmion Logic Gates Based on Skyrmion Tunneling," *Phys Rev Appl*, vol. 17, no. 6, p. 064035, Jun. 2022, doi: 10.1103/PhysRevApplied.17.064035.

[47] D. Pinna *et al.*, "Skyrmion Gas Manipulation for Probabilistic Computing," *Phys Rev Appl*, vol. 9, no. 6, Jun. 2018, doi: 10.1103/PhysRevApplied.9.064018.

[48] D. Kechrakos *et al.*, "Skyrmions in nanorings: A versatile platform for skyrmionics," *Phys Rev Appl*, vol. 20, no. 4, p. 044039, Oct. 2023, doi: 10.1103/PhysRevApplied.20.044039.

[49] N. E. Penthorn, X. Hao, Z. Wang, Y. Huai, and H. W. Jiang, "Experimental Observation of Single Skyrmion Signatures in a Magnetic Tunnel Junction," *Phys Rev Lett*, vol. 122, no. 25, p. 257201, Jun. 2019, doi: 10.1103/PhysRevLett.122.257201.

[50] X. Zhang *et al.*, "Skyrmions in Magnetic Tunnel Junctions," *ACS Appl Mater Interfaces*, vol. 10, no. 19, pp. 16887–16892, May 2018, doi: 10.1021/acsami.8b03812.

[51] J.-V. Kim, F. Garcia-Sanchez, J. Sampaio, C. Moreau-Luchaire, V. Cros, and A. Fert, "Breathing modes of confined skyrmions in ultrathin magnetic dots," *Phys Rev B*, vol. 90, no. 6, p. 064410, Aug. 2014, doi: 10.1103/PhysRevB.90.064410.

[52] Z. X. Liu and H. Xiong, "Ultra-slow spin waves propagation based on skyrmion breathing," *New J Phys*, vol. 25, no. 10, Oct. 2023, doi: 10.1088/1367-2630/ad05a5.

[53] C. E. A. Barker, E. Haltz, T. A. Moore, and C. H. Marrows, "Breathing modes of skyrmion strings in a synthetic antiferromagnet multilayer," *J Appl Phys*, vol. 133, no. 11, Mar. 2023, doi: 10.1063/5.0142772.

[54] S. Komineas and P. E. Roy, "Breathing skyrmions in chiral antiferromagnets," *Phys Rev Res*, vol. 4, no. 3, Jul. 2022, doi: 10.1103/PhysRevResearch.4.033132.

[55] B. F. McKeever, D. R. Rodrigues, D. Pinna, A. Abanov, J. Sinova, and K. Everschor-Sitte, "Characterizing breathing dynamics of magnetic skyrmions and antiskyrmions within the Hamiltonian formalism," *Phys Rev B*, vol. 99, no. 5, p. 054430, Feb. 2019, doi: 10.1103/PhysRevB.99.054430.

[56] N. E. Penthorn, X. Hao, Z. Wang, Y. Huai, and H. W. Jiang, "Experimental Observation of Single Skyrmion Signatures in a Magnetic Tunnel Junction," *Phys Rev Lett*, vol. 122, no. 25, p. 257201, Jun. 2019, doi: 10.1103/PhysRevLett.122.257201.

[57] C. Zhang *et al.*, "Spin current pumped by confined breathing skyrmion," *New J Phys*, vol. 22, no. 5, p. 053029, May 2020, doi: 10.1088/1367-2630/ab83d6.

[58] R. Tomasello *et al.*, "Field-driven collapsing dynamics of skyrmions in magnetic multilayers," *Phys Rev B*, vol. 107, no. 18, p. 184416, May 2023, doi: 10.1103/PhysRevB.107.184416.

[59] O. Lee *et al.*, "Tunable gigahertz dynamics of low-temperature skyrmion lattice in a chiral magnet," *Journal of Physics Condensed Matter*, vol. 34, no. 9, Mar. 2022, doi: 10.1088/1361-648X/ac3e1c.

[60] M. Lonsky and A. Hoffmann, "Coupled skyrmion breathing modes in synthetic ferri- And antiferromagnets," *Phys Rev B*, vol. 102, no. 10, Sep. 2020, doi: 10.1103/PhysRevB.102.104403.

[61] Y. Onose, Y. Okamura, S. Seki, S. Ishiwata, and Y. Tokura, "Observation of magnetic excitations of skyrmion crystal in a helimagnetic insulator Cu 2OSeO 3," *Phys Rev Lett*, vol. 109, no. 3, Jul. 2012, doi: 10.1103/PhysRevLett.109.037603.

[62] F. Büttner, I. Lemesh, and G. S. D. Beach, "Theory of isolated magnetic skyrmions: From fundamentals to room temperature applications," *Sci Rep*, vol. 8, no. 1, p. 4464, Mar. 2018, doi: 10.1038/s41598-018-22242-8.

[63] A. Sengupta, C. M. Liyanagedera, B. Jung, and K. Roy, "Magnetic Tunnel Junction as an On-Chip Temperature Sensor," *Sci Rep*, vol. 7, no. 1, p. 11764, Sep. 2017, doi: 10.1038/s41598-017-11476-7.

[64] A. G. Qoutb and E. G. Friedman, "Distributed Spintronic/CMOS Sensor Network for Thermal-Aware Systems," *IEEE Trans Very Large Scale Integr VLSI Syst*, vol. 28, no. 6, pp. 1505–1512, Jun. 2020, doi: 10.1109/TVLSI.2020.2981443.

[65] Xiaobin Wang, Yiran Chen, Ying Gu, and Hai Li, "Spintronic Memristor Temperature Sensor," *IEEE Electron Device Letters*, vol. 31, no. 1, pp. 20–22, Jan. 2010, doi: 10.1109/LED.2009.2035643.

[66] J. Urrestarazu Larrañaga *et al.*, "Electrical Detection and Nucleation of a Magnetic Skyrmion in a Magnetic Tunnel Junction Observed via Operando Magnetic Microscopy," *Nano Lett*, vol. 24, no. 12, pp. 3557–3565, Mar. 2024, doi: 10.1021/acs.nanolett.4c00316.

[67] D. Jiang *et al.*, "Substrate-induced spin-torque-like signal in spin-torque ferromagnetic resonance measurement," *Phys Rev Appl*, vol. 21, no. 2, p. 024021, Feb. 2024, doi: 10.1103/PhysRevApplied.21.024021.

[68] A. Giordano, G. Finocchio, L. Torres, M. Carpentieri, and B. Azzerboni, "Semi-implicit integration scheme for Landau–Lifshitz–Gilbert-Slonczewski equation," *J Appl Phys*, vol. 111, no. 7, p. 07D112, Apr. 2012, doi: 10.1063/1.3673428.

[69] A. Soumyanarayanan, N. Reyren, A. Fert, and C. Panagopoulos, "Emergent phenomena induced by spin–orbit coupling at surfaces and interfaces," *Nature*, vol. 539, no. 7630, pp. 509–517, Nov. 2016, doi: 10.1038/nature19820.

[70] E. Grimaldi et al., "Single-shot dynamics of spin–orbit torque and spin transfer torque switching in three-terminal magnetic tunnel junctions," *Nat Nanotechnol*, vol. 15, no. 2, pp. 111–117, 2020, doi: 10.1038/s41565-019-0607-7.

[71] K. M. Lee, J. W. Choi, J. Sok, and B. C. Min, "Temperature dependence of the interfacial magnetic anisotropy in W/CoFeB/MgO," *AIP Adv*, vol. 7, no. 6, pp. 1–8, 2017, doi: 10.1063/1.4985720.

[72] R. Tomasello et al., "Origin of temperature and field dependence of magnetic skyrmion size in ultrathin nanodots," *Phys Rev B*, vol. 97, no. 6, p. 060402, Feb. 2018, doi: 10.1103/PhysRevB.97.060402.

[73] L. Rózsa, U. Atxitia, and U. Nowak, "Temperature scaling of the Dzyaloshinsky-Moriya interaction in the spin wave spectrum," *Phys Rev B*, vol. 96, no. 9, pp. 1–11, 2017, doi: 10.1103/PhysRevB.96.094436.

[74] R. Moreno, R. F. L. Evans, S. Khmelevskyi, M. C. Muñoz, R. W. Chantrell, and O. Chubykalo-Fesenko, "Temperature-dependent exchange stiffness and domain wall width in Co," *Phys Rev B*, vol. 94, no. 10, pp. 1–6, 2016, doi: 10.1103/PhysRevB.94.104433.

[75] A. Ghosh, J. F. Sierra, S. Auffret, U. Ebels, and W. E. Bailey, "Dependence of nonlocal Gilbert damping on the ferromagnetic layer type in ferromagnet/Cu/Pt heterostructures," *Appl Phys Lett*, vol. 98, no. 5, Jan. 2011, doi: 10.1063/1.3551729.

[76] Y. Zhao et al., "Experimental Investigation of Temperature-Dependent Gilbert Damping in Permalloy Thin Films," *Sci Rep*, vol. 6, Mar. 2016, doi: 10.1038/srep22890.

[77] H. Maier-Flaig et al., "Temperature-dependent magnetic damping of yttrium iron garnet spheres," *Phys Rev B*, vol. 95, no. 21, Jun. 2017, doi: 10.1103/PhysRevB.95.214423.

[78] V. P. Kravchuk, D. D. Sheka, U. K. Rößler, J. van den Brink, and Y. Gaididei, "Spin eigenmodes of magnetic skyrmions and the problem of the effective skyrmion mass," *Phys Rev B*, vol. 97, no. 6, p. 064403, Feb. 2018, doi: 10.1103/PhysRevB.97.064403.

[79] N. K. Duong et al., "Magnetization reversal signatures of hybrid and pure Néel skyrmions in thin film multilayers," *APL Mater*, vol. 8, no. 11, p. 111112, Nov. 2020, doi: 10.1063/5.0022033.

[80] P. N. Skirdkov and K. A. Zvezdin, "Spin-Torque Diodes: From Fundamental Research to Applications," *Ann Phys*, vol. 532, no. 6, Jun. 2020, doi: 10.1002/andp.201900460.

[81] J. Zhang, M. Bachman, M. Czerner, and C. Heiliger, "Thermal Transport and Nonequilibrium Temperature Drop Across a Magnetic Tunnel Junction," *Phys Rev Lett*, vol. 115, no. 3, Jul. 2015, doi: 10.1103/PhysRevLett.115.037203.

[82] H. Jang, L. Marnitz, T. Huebner, J. Kimling, T. Kuschel, and D. G. Cahill, "Thermal Conductivity of Oxide Tunnel Barriers in Magnetic Tunnel Junctions Measured by Ultrafast Thermoreflectance and Magneto-Optic Kerr Effect Thermometry," *Phys Rev Appl*, vol. 13, no. 2, Feb. 2020, doi: 10.1103/PhysRevApplied.13.024007.

[83] B. Fang et al., "Giant spin-torque diode sensitivity in the absence of bias magnetic field," *Nat Commun*, vol. 7, Apr. 2016, doi: 10.1038/ncomms11259.